Authors:
Roee Gilron[1], Jonathan Rosenblatt[2], Oluwasanmi Koyejo[3], Russell A. Poldrack[3], Roy Mukamel[1]

Institutions:
[1] Sagol School of Neuroscience; School of Psychological Sciences, Tel Aviv University, Tel Aviv, Israel, [2] Department of Industrial Engineering and Management; Zlotowsky Center for Neuroscience Ben Gurion University, Beer Sheva, Israel, [3] Department of Psychology, Stanford University, Stanford, CA

Abstract:

Multivoxel pattern analysis (MVPA) has gained enormous popularity in the neuroimaging community over the past few years. At the group level, most MVPA studies adopt an "information based" approach in which the sign of the effect of individual subjects is discarded and a non-directional summary statistic is carried over to the second level. This is in contrast to a directional "activation based" approach typical in univariate group level analysis, in which both signal magnitude and sign are taken into account. The transition from examining effects in one voxel at a time vs. several voxels (univariate vs. multivariate) has thus tacitly entailed a transition from directional to non-directional signal definition at the group level. While a directional group-level MVPA approach implies that individuals have similar multivariate spatial patterns of activity, in a non-directional approach each individual may have a distinct spatial pattern. Using an experimental dataset, we show that directional and non-directional group-level MVPA approaches uncover distinct brain regions with only partial overlap. We propose a method to quantify the degree of spatial similarity in activation patterns over subjects. Applied to an auditory task, we find higher values in auditory regions compared to control regions.

*Intro:*

In the last decade, the use of multivoxel pattern analysis (MVPA) to analyze fMRI data has grown substantially and is now commonplace (Haxby, 2012; Haynes and Rees, 2006; Kriegeskorte et al., 2006a; Poldrack and Farah, 2015; Tong and Pratte, 2012). The increasing use of MVPA approaches compared to classical univariate approaches has also tacitly implied a move from a non-directional to a directional definition of signal at the group level. Here we expose this shift in the definition of signal, impacting popular MVPA approaches in group inference. In addition, we suggest a novel application of recently developed statistical measures to address this issue. Our proposed statistic has the added benefit of quantifying the degree to which subjects share multivariate patterns of activity at the group level.

We focus on examples in which the signal of two conditions is compared. In a typical mass-univariate analysis, the BOLD signal in each individual voxel is examined separately by comparing values between conditions at the individual subject level (first level). This is typically conducted by performing a t-test examining the null hypothesis that the expected response is not different across conditions. In multivariate approaches, a spatial pattern of activity is compared (Haxby et al., 2001). Commonly in such cases, supervised machine learning approaches such as linear discriminant analysis or support vector machines (Kragel et al., 2012; Misaki et al., 2010; Mur et al., 2009; Tong and Pratte, 2012) are used, and their results are compared against an empirical null distribution - putatively centered around chance classification levels.

At the second (group) level, univariate studies use a random effects (RFX) analysis to examine whether the average difference between two conditions is consistent across subjects. If the mean difference between conditions is significantly different from zero (as examined using a t-test for example), the voxel is declared responsive at the group level. Since the difference between conditions is signed, to reject the null one must show a *directional* group-wise effect (Fig. 1A). A directional effect is one in which most subjects display a consistent (either positive or negative) effect in a given voxel. This takes into account both magnitude and sign (direction) of the effect. This directional effect has been termed "activation based" to emphasize its origin. If, for example, we had a cohort of subjects in which half of the sample showed an *increase* in their response to one condition relative to the other while the other half showed a *decrease* of equal magnitude in their response – a second level directional analysis would not define such a group effect as signal. A *directional* group wise effect implies that subjects share a similar spatial pattern of activity, henceforth referred to as *similarity.* Put differently, variability in pattern similarity is part of the RFX null hypothesis and not part of the alternative. Although there is a strong effect size at the individual subject level, at the group level there is no significant effect under such a directional definition of signal. Indeed, a *directional* approach is the commonly adopted signal definition in second-level mass-univariate RFX analysis.

# Univariate & Multivariate Signal Definition

## A  Univariate Signal - Group Level

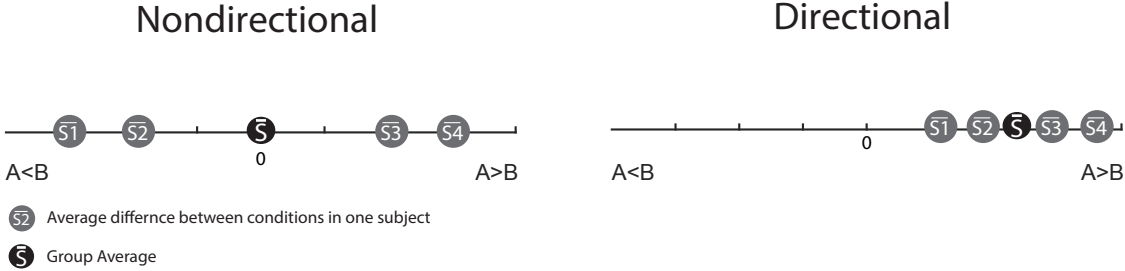

## B  Multivariate Signal - Group Level

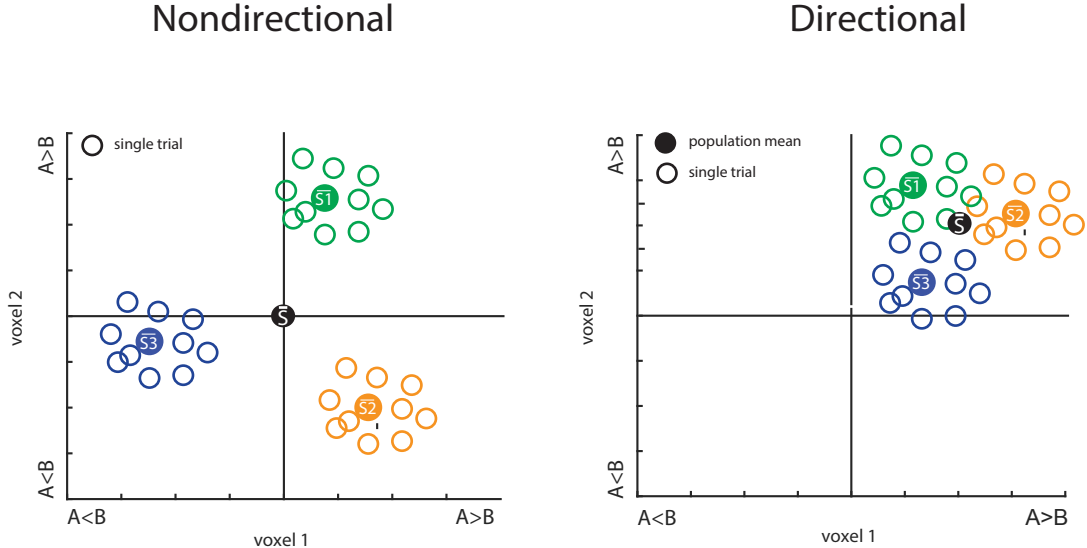

Figure 1 Caption:

Univariate and multivariate signal definition. This schematic diagram represents the different signal definitions in univariate and multivariate approaches employing either a directional or non-directional analysis. (A) Univariate group level analysis. Grey colored circles represent the average difference (contrast) between conditions of interest (A and B) of individual subjects. The group average is represented by a filled black circle. In a directional univariate analysis, activation is defined as a group average that is different from zero (conceptual example - top right). In contrast, in a non-directional univariate analysis the voxel may be declared active even if the mean of the contrast across subjects is zero (top left). (B) Multivariate group level analysis. Empty circles represent single trials, filled circles represent average difference of single subject, and black filled circle is group average. In a non-directional multivariate analysis a beam is considered active provided that subjects are not all at zero (left). Note that the group average can be centered at zero. In contrast, in a directional multivariate analysis subjects share a spatial pattern of activity such that the beam is considered active if the group average is away from zero (conceptual example - right). The non-directional approach is the most commonly used in the $2^{nd}$ level multivariate analysis, whereas the directional approach is the most commonly used in $2^{nd}$ level univariate analysis.

In contrast, the large majority of MVPA studies to date have adopted a *non-directional* (information based) definition of signal at the group level (Fig. 1B). In a *non-directional* analysis, a certain statistic (usually classification accuracy) is calculated at the individual subject level, and this statistic is then carried over to the second level. Note that as opposed to the t-statistic (or beta contrast), the accuracy statistic is directionless, thus the sign of the effect at the first level is lost and only its magnitude is passed on to the second level. In the example described earlier (see also Fig 1A – left) half of the subjects show an increase in their response to condition 1 vs. condition 2 while the other half of subjects show a decrease of equal magnitude. Thus effect size at the individual subject level is large and would be reflected in a corresponding high statistical value (e.g. classification accuracy) that is carried to the second level. Since all subjects have a large effect size, such a case would be detected by a non-directional $2^{nd}$ level analysis, irrespective of the fact that different subjects show completely opposite patterns of responses. The equivalent univariate null hypothesis of a *non-directional* signal definition is that across subjects, the expected *absolute value* of the effect in a given voxel is zero. Thus a signal would be detected even if some subjects show a positive effect and others show a negative effect. This non-directional univariate approach is seldom taken when conducting group-level analysis since the biological significance of such an effect would be deemed suspect. That is, it would be challenging to interpret a study in which half of the subjects show an increase in BOLD response whereas the other half show a decrease in BOLD response in a given voxel to a given contrast. Thus the transition from examining effects in one voxel at a time vs. several voxels (univariate vs. multivariate) has tacitly entailed a transition from directional to non-directional signal definition at the 2$^{nd}$ level. Studies that opted for an MVPA *directional* signal definition are rare, primarily occurring in cases where classifiers are used to predict new individuals (Helfinstein et al., 2014).

Multivariate non-directional 2$^{nd}$ level analysis implies a fundamentally different *definition of signal* compared to the traditional univariate 2$^{nd}$ level directional analysis. This represents an implicit paradigm shift the field has undergone. There is no a-priori reason to believe that moving from univariate (single voxel) to multivariate analysis (2 voxels or more) requires a redefinition of the null and alternative hypothesis in signal definition. Moreover the original motivation of multivariate approaches was to uncover weak distributed signals as well as information at finer spatial scales than fMRI affords (Haxby, 2012; Haxby et al., 2001; Haynes and Rees, 2006; Kamitani and Tong, 2005; Kriegeskorte et al., 2006b). These papers did not make any explicit hypothesis about the differing nature of the signal across subjects. The practice of carrying over an unsigned statistical measure to second level analysis is subject to a variety of confounds controlled for in a traditional univariate directional analysis (Todd et al., 2013). These confounds provide alternative explanations primarily for the results of non-directional multivariate approaches so a careful definition of the type of signal one expects to find can contribute to minimizing errors.

Using an empirical data set we show that divergent definitions of the null hypothesis governing 2$^{nd}$-level *directional* and *non-directional* analysis yield different results. Furthermore, we suggest a directional statistic that allows one to quantify the degree of similarity between subjects' patterns of activity. We compare this to a number of other potential metrics in Appendix 1.

*Material and Methods:*

Data Set:

We used an fMRI dataset of a localizer used to identify areas sensitive to human voices (vs. non-man-made sounds) in the auditory cortex. The experimental procedure is described in detail in their original paper (Pernet et al., 2015), therefore we describe it here in brief. The voice localizer is a 10:20 min block design fMRI experiment. The experiment consisted of 40 blocks, each lasting 8 seconds, of human vocalizations (20 blocks) or non-vocal (20 blocks) stimuli. A few periods of silence (10 seconds) were interspersed between the experimental blocks to allow the hemodynamic response to relax. Vocal blocks were primarily sounds of human vocal origin obtained from 47 speakers while non-vocal sounds were mostly from natural or man-made sources (like cars). Scans were acquired using a 3T Siemens (Erlangen, Germany) Tim Trio using a repetition time (TR) of 2s, an echo time (TE) of 30ms, and 3x3x3.3mm resolution. Additional scan parameters can be found in the original paper. Data were graciously shared by the authors and can be found at https://openfmri.org/dataset/ds000158.

fMRI Pre-processing:

We used the pre-processing analysis code used by Pernet and colleagues which can be found on the OpenfMRI link above. The original data set contained 218 subjects, but since the rostral part of the frontal cortex was not scanned in some of the subjects, whole brain functional coverage was available for 150 subjects from which we randomly chose a subset of 20 subjects for our analysis. We chose 20 subjects in order to obtain a sample size concordant with many fMRI studies and to avoid trivial power gains. Data were analyzed using SPM12b (r6225 – Welcome Department of Cognitive Neurology, University College London). Pre-processing consisted of slice time correction, motion correction (6 parameters), co-registration of the structural image to the mean functional image and normalization of the structural image to the Montreal Neurological Institute (MNI) space (diffeomorphic normalization with the forward deformation field computed during segmentation, data was resampled at 2mm isotropic with 4$^{th}$ degree B-spline interpolation). These spatial transformations were then applied to the functional images to achieve normalization to the Montreal Neurological Institute (MNI) space. Data were high pass filtered (1/128s) to remove slow drift.

In accordance with the standard MVPA pipeline, we used a design matrix that contained separate regressors for each trial. Each regressor was modelled by convolving a boxcar function describing the timing of stimulus events with the canonical hemodynamic response function (HRF) used in SPM. Since strong correlations between trial-wise beta estimates still existed in the data after the use of SPM12's default AR(1) serial correlation model, we used SPM12's AR(6) serial correlation model to remove correlations in the beta estimates (see(Gilron et al., 2016) and statistical significance section below). A separate beta value was estimated for each block resulting in a total of 40 beta values per subject (20 vocal, 20 non-vocal). Since scan coverage was not identical across subjects, we created a Boolean 'AND' map of all subjects' functional data masks in order to allow us to easily compare only signals in voxels that are common across all subjects. This resulted in a matrix of 40 beta values X 32,482 voxels per subject. In both the directional and non-directional analysis detailed below, we used a searchlight approach similar to the one employed in a previous paper (Krasovsky et al., 2014). For each center voxel, beta values from its 26 closest voxels were used in the data analysis. Thus each searchlight beam in a single subject was represented by a 40 x 27 matrix corresponding to 40 beta values per voxel (20 vocal, 20 non-vocal trials) by 27 voxels (center voxel + 26 closest neighbors using Euclidian distance).

Detecting signal in a searchlight beam – the statistical test

When facing a multivariate comparison between two conditions, most neuroimaging studies have employed a supervised machine learning approach in which performance is assessed through testing of out-of-sample generalization (e.g., via the cross validated prediction accuracy). While this approach is useful in assessing the generalizability of the results, for the mere purpose of localization (i.e. where in the brain are there significant differences between conditions) it is substantially more conservative than population tests. A number of studies have suggested the use of in-sample hypothesis testing over out-of-sample classification for multivariate comparison (Allefeld and Haynes, 2014; Kriegeskorte et al., 2006a). There is a wide body of statistical literature concerned with detecting multivariate differences between populations (Anderson, 2003). Allefeld and Haynes (2014) proposed a variation on Hotelling's Trace as their multivariate test. Here we use a related multivariate statistical test which is better equipped to deal with cases in which the number of features (voxels in the searchlight) is larger than the number of observations (trials or subjects in our case). This test, developed by Strivastava and Du (2008), is both quick to compute, and more powerful than Hotteling type tests for the dimensions of a searchlight used in a typical MVPA fMRI setup.

Both the directional and non-directional tests have the same general structure, which consists of testing for the expectation of the group effect based on some first (subject) level summary statistic. We thus denote $T_i$ as subject *i*'s summary statistic of a beam centered at voxel $v$ (voxel index omitted). We denote by $T$ the group level summary of the same beam. Under the summary statistic approach, $T = (T_1, \ldots, T_n)$, where $n$ is the number of subjects. We will also denote by $p$ the number of voxels in a beam, and by $n$ the number of repeats of each stimulus (trials), which is the same for the two stimuli in our balanced design.

In the most general case, each beam is fitted with a multivariate general linear model, and then signal detection can be performed with any test for the coefficients of a MANOVA such as Wilk's Lambda, Pillai-Bartlett Trace, Lawley-Hotelling Trace, or Roy's Greatest Root test (Anderson, 2003). This was indeed the framework in Allefeld and Haynes (2014). In our two-stimuli case, all these tests collapse to the classic Hotelling test, which is perhaps the best known multivariate test. It is however notoriously low powered when the number of parameters ($p$) is in the same order as the number of samples (number of subjects or trials) (Dempster, 1963). In our analysis, we found the Srivastava-Du (2008) statistic to be the most powerful metric for search-light MVPA when compared to Hotelling's, and Dempster's statistic (1963). We also conjecture that the Schaffer-Strimmer statistics (Schafer and Strimmer, 2005) should have similar performance, but this has not been tested.

We now present our multivariate directional and non-directional tests. In order to assess statistical significance we use the permutation scheme of Stelzer et al. (2013) as discussed in the "Significance Testing" section.

Directional analysis at the group level

Our statistic for detecting directional signal consists of applying the one-sample multivariate test described in Strivastava and Du (2008) to the directional summary from the first level. Formally, let $c_i$ be subject $i$'s vector valued estimated contrast of interest. In our example, each of the $p$ coordinates of $c_i$ encodes the difference in the mean response between vocal and non-vocal response. More generally, it may be the output of any contrast in a multivariate linear model.

The directional test we propose consists of the following two levels:

$$T_i^D := c_i, \tag{1}$$

$$T^D := T^{2008}(T_1^D, \ldots, T_n^D) = T^{2008}(c_1, \ldots, c_n), \tag{2}$$

where $T^{2008}(\ldots)$ is the one-sample Srivastava-Du test, which is defined as

$$T^{2008}(c_i, \ldots, c_n) := \frac{n\bar{c}^t D^{-1} \bar{c} - \frac{n^* p}{n^* - 2}}{\sqrt{2d(tr(R^2) - \frac{p^2}{n^*})}}, \tag{3}$$

where $n^* = n - 1$; $\bar{c} := 1/n \sum_i c_i$; $S := 1/n^* \sum_i (c_i - \bar{c})(c_i - \bar{c})^t$; $D := diag(S)$; $R := D^{-1/2} S D^{-1/2}$; and $d := 1 + \frac{tr(R^2)}{p^{3/2}}$.

In this test each subject is essentially summarized by its raw contrast estimate. The first level statistic, $c_i$, is trivially directional. The test statistic $T^{2008}$ can be seen as Hotelling's $T^2$ computed under a spatial (between voxel) independence assumption, and then corrected to relax this assumption. For more on the design and motivation of this test statistic, see Strivastava and Du (2008).

Non-Directional analysis at the group level

For the non-directional version of the group test, each subject is summarized by a non-directional measure of signal. Hotelling's two group test is a natural candidate, but again, we will want to replace it with a high-dimensional version in which the number of features (voxels) can be larger than the number of samples (trials in our case).

Seeing the two conditions in our example as a balanced block design so that $x_{i,j}$ denotes subject *i*'s *j*'th response to a vocal stimulus, and $y_{i,j}$ the same for a non vocal stimulus, the non-directional test we propose has the following form:

$$T_i^{ND} := T^{2013}(x_1, \ldots, x_m, y_1, \ldots, y_m), \tag{4}$$

$$T^{ND} := \frac{1}{n}\sum_{i=1}^{n} T_i^{ND}, \tag{5}$$

where $T^{2013}$ is the two-sample Srivastava-Du test defined as

$$T^{2013}(x_1, \ldots x_m, y_1, \ldots y_m) := \frac{\frac{m}{2}\bar{\delta}^t D^{-1}\bar{\delta} - p}{\sqrt{2d(tr(R^2) - \frac{p^2}{m^*})}}, \tag{6}$$

where $m^* = m - 1$; $\bar{x} := \frac{1}{m}\sum_j x_j$ ; $\bar{y} := \frac{1}{m}\sum_j y_j$ ; $\bar{\delta} := \bar{x} - \bar{y}$; $S_x := 1/m^* \sum_i(x_j-\bar{x})(x_j - \bar{x})^t$; $S_y := \frac{1}{m^*}\sum_j(y_j-\bar{y})(y_j - \bar{y})^t$; $S := (S_x + S_y)/2$ ; $D := diag(S)$, $R := D^{-1/2}SD^{-1/2}$; and $d := 1 + \frac{tr(R^2)}{p^{3/2}}$.

Like the single sample case, this test can be seen as a two sample Hotelling $T^2$ test, corrected for the relaxation of the assumption of (spatial) independence.

In this test each subject is summarized by a beam-wise statistic, in this case $T^{2013}$, which is later averaged over subjects. To verify that the first level statistic is non-directional, one may is to observe that $T^{2013}$ is a scaled and shifted quadratic form in the difference between group means ($\bar{\delta}$). As such, and just like the squared univariate $t$-statistic, it grows when $\bar{x} > \bar{y}$, and also when $\bar{x} < \bar{y}$. We also note that while we assumed a balanced design, this assumption is relaxed in (Srivastava et al., 2013) Section 4.4.

# Analysis Scheme

## 2nd level Nondirectional MVPA

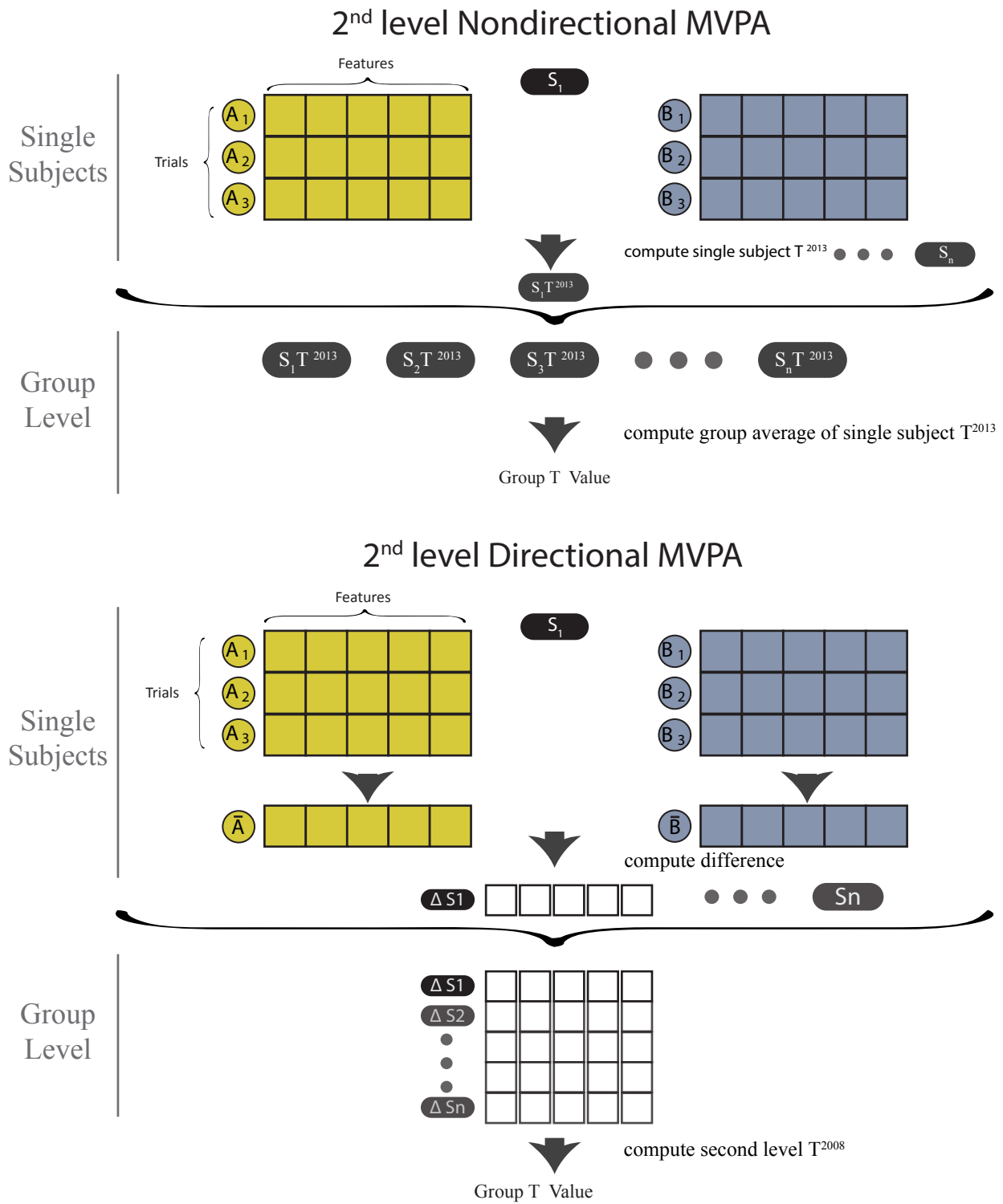

Figure 2 Caption:

The top panel describes the analysis scheme for non-directional MVPA. At the first level, for each center voxel, in each subject, a matrix (trials x voxels) from each condition is used in order to calculate $T^{2013}$. The circles represent trial labels and the squares activity in a particular voxel (feature). Each row represents a particular trial. The $T^{2013}$ value is non-signed and is calculated for each center voxel for each subject. On the second level, the single subject $T^{2013}$ values are averaged to create a group $T^{ND}$ map composed of average $T^{2013}$ across subjects for each voxel.

The bottom panel describes the analysis scheme for directional MVPA. Here the first level summary statistic of each subject ($T_i^D$) is simply the difference between the average activity in each condition. At the second level, this signed summary statistic is aggregated across subjects and the group $T^D$ value is calculated using the $T^{2008}$ statistic for each center voxel.

Classifier based version of directional and non-directional test

To demonstrate that the distinction between directional and non-directional effects is not restricted to multivariate testing, but also applies to the more common classification approach, we now introduce the classification versions of directional and non-directional analysis. We employ the same analysis scheme described above for directional as well as non-directional tests, using a 5-fold stratified cross validation procedure. In the non-directional case, we computed the cross validated classification accuracy of a linear SVM, within-subject—within-searchlight. At the second level, accuracy was averaged across subjects. In the directional case, cross validation was employed across subjects instead of across trials within subject. We used the average within-subject contrast as input to a group level SVM. A holdout of subjects was used to cross validate the accuracy. The main purpose of this analysis was not to showcase any power differences between classification and testing approaches but to test whether classification based approaches are also sensitive to directional and non-directional definitions of signal. For power comparisons between the testing and classifcation approach see Rosenblatt et. al. (2016).

Significance Testing

To threshold our directional and non-directional group level analysis we employ the same non-parametric permutation scheme described by Stelzer et al. (2013). We shuffle the condition labels across trials within each subject, and compute $T^D$ and $T^{ND}$ value using the same pipeline described above to generate a shuffled map. For each shuffled permutation map we use the same shuffling scheme across all searchlight beams so that spatial correlations in the noise are conserved. It is important to note that this dataset had strong correlations between trial-wise beta estimates before the use of the AR(6) model to whiten the noise process. If one were to use the default AR(1) model in SPM, a naïve permutation scheme would underestimate the number of significant voxels due to dependence between trial-wise estimates (see Gilron et. al. 2016 for further details and a review of the problems associated with correlations between parameter estimates and non-parametric significance testing). Once we used an AR(6) model to whiten the noise, trial-wise estimates were no longer correlated and in accordance with Stelzer (2013), we computed 5,000 shuffled label whole brain searchlight maps for each subject. We created group level shuffled-label maps by averaging randomly selected maps from each subject's shuffled maps (with replacement). Within each voxel we used the distribution of shuffled values to compute a corresponding voxel-wise p-value for both the $T^{ND}$ and $T^D$ maps. In this way we associate a p-value with each searchlight beam in the brain and can submit these p-values to false discovery rate (FDR) control with the BH procedure (Benjamini and Hochberg, 1995) to create a binary map of the center voxels which pass significance. An example of this implementation can be found in the accompanying code.

*Results:*

Using whole brain directional MVPA searchlight analysis we found a number of significant regions (voxelwise FDR ≤ 0.05) including bilateral primary and secondary auditory cortices, right precuneus and left angular gyrus (see Table 1 in the appendix for the full list). A total of 1,376 voxels passed significance (Fig 3. red and blue). Our whole brain MVPA non-directional searchlight analysis revealed a number of partially overlapping brain regions which survived FDR control (FDR ≤ 0.05) including bilateral primary auditory cortex, and bilateral amygdala. A total of 1,331 voxels passed significance (Fig. 3, green and blue). Overlapping voxels that were detected in both the directional and non-directional analysis are shown in blue (a total of 658 voxels). A table detailing all clusters can be found in the appendix.

# Signal detection using directional and non-directional testing

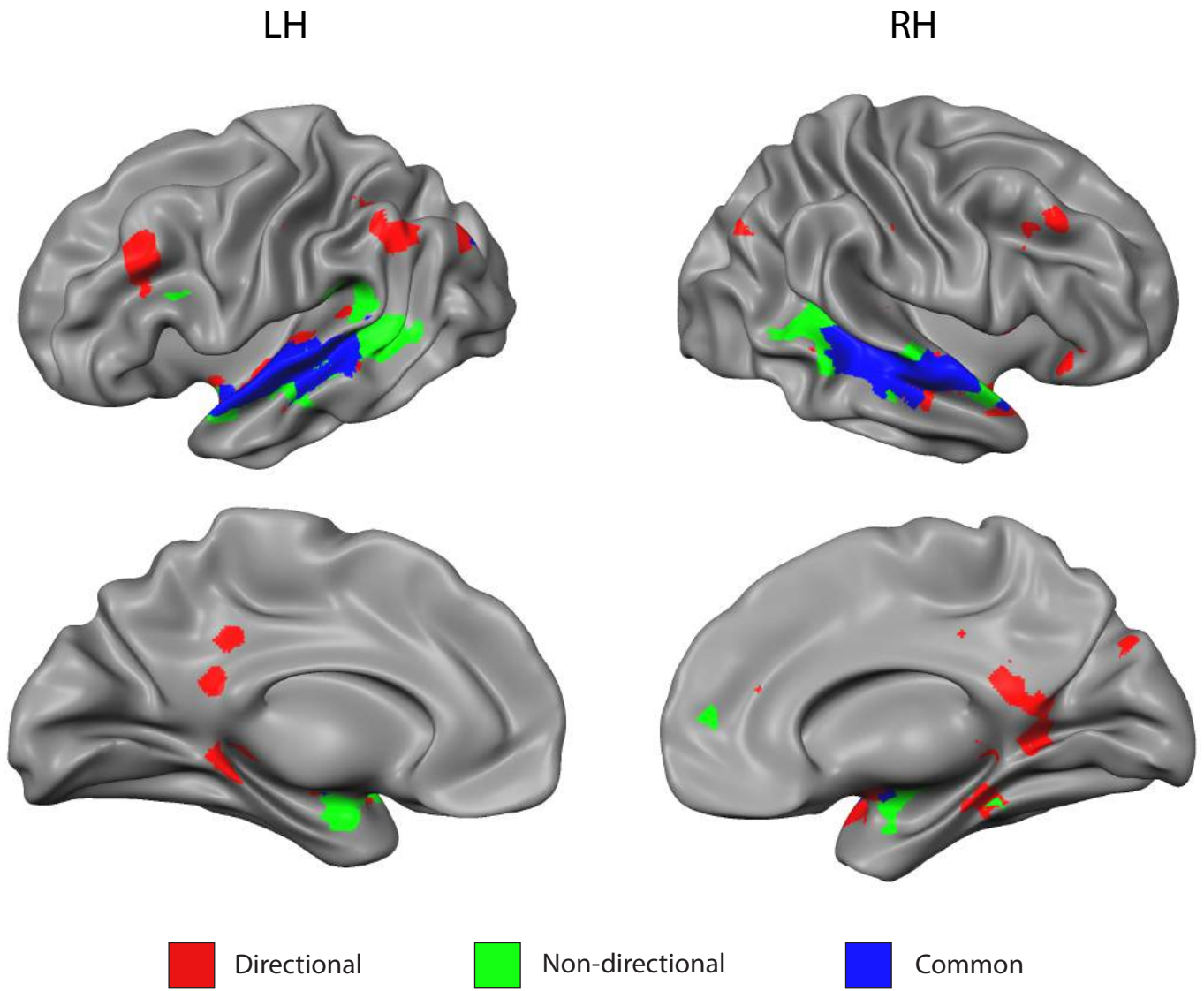

Figure 3 Caption:

Signal detection using directional and non-directional testing. These binary maps represent an overlay of voxels which were declared significant using only a directional analysis (red), only a non-directional analysis (green), or both (blue).

Our directional multivariate statistic not only detects regions in which the subjects share spatial patterns of activity, but also quantifies their degree of similarity. Higher $T^D$ values correspond to a larger degree of similarity across subjects. To showcase this phenomena we overlaid the Boolean maps from the directional analysis seen in Figure 3 (blue and red) with their actual directional $T^D$ values (Fig. 4a). Note the spatial gradient in $T^D$ values demonstrating that spatial agreement between the 20 subjects in our dataset smoothly decays from the human voice areas identified by Pernet et. al. (2015) along the superior temporal gyrus, to its boundaries in the sulci. This highlights the potential of $T^D$ to quantify the degree of similarity between subjects' spatial pattern of activity. For comparison see figure 3 in Pernet et. al. (2015) showing a probability map of 218 subjects. For additional potential measures of spatial similarity we examined see Appendix 1.

To verify that the spatial similarity between subjects is driven by evoked signals, and not noise, we plotted $T^D$ values in a number of predefined anatomical regions (Fig. 4b). For instance, we expect low $T^D$ values in control regions such as the ventricles and white matter in which no spatial activation similarity across subjects is expected.

# Quantifing Pattern Overlap Across Subjects

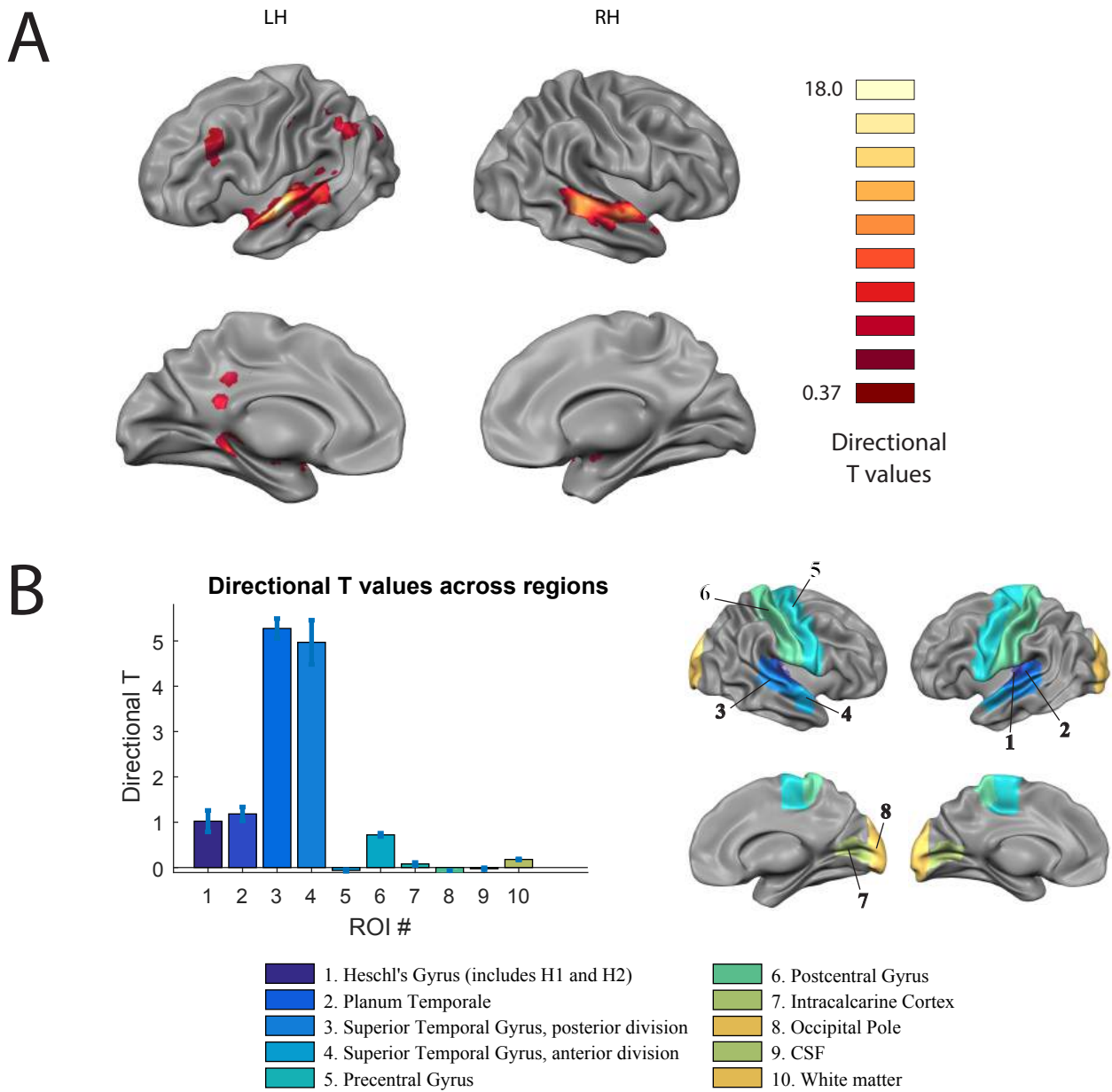

Figure 4 Caption:

A) Directional T values overlaid on significant voxels defined using directional analysis (red and blue voxels in figure 3). B) Mean directional T values across voxels in several pre-defined anatomical regions selected using the Harvard-Oxford atlas. Note the high values in auditory regions and low values in control regions. Further note the gradient of values between STG (regions 3 & 4; high order auditory regions) and primary auditory regions (regions 1 & 2).

We also compared our directional/non-directional testing-based results to their equivalent classification-based results in this dataset. As expected classification-based approaches (using linear SVM) are also sensitive to directional and non-directional definitions of signal. In a similar fashion to the testing approach, a classification based approach uncovers distinct brain regions with partial overlap depending on the use of a directional or non-directional analysis. In our classification based directional analysis (Fig 5. - red and blue), a total of 271 voxels survived FDR control (FDR ≤ 0.05). In the non-directional classification based analysis a total of 441 voxels survived FDR control (FDR ≤ 0.05). Overlapping voxels common to both the directional and non-directional analysis (shown in blue) were also found (147 voxels). Our analysis is not designed to examine power differences between classification and testing. For a rigorous comparison showcasing the power advantages of testing over classification see Rosenblatt et. al. (2016). In agreement with the power advantage of testing vs. classification simulation results reported by Rosenblatt et. al. (2016), in the directional analysis our testing-based approach uncovered 72.7% of the voxels found by a classification based approach whereas the classification approach only uncovered 14.3% of the voxels found using testing. In the non-directional analysis our testing based approach uncovered 90.5% of the voxels found by classification whereas the classification approach only uncovered 30.0% of the voxels found using testing.

# Signal detection using directional and non-directional classification

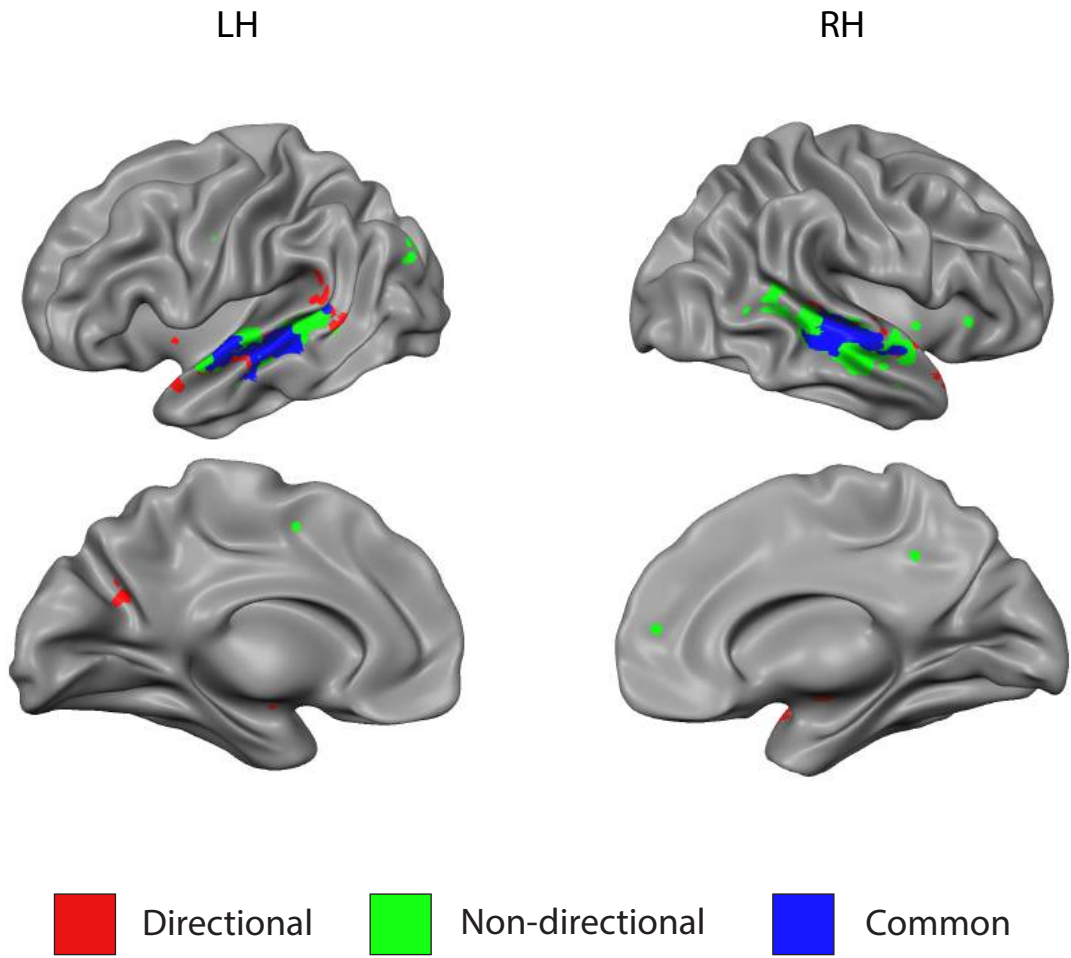

Figure 5 Caption:

Signal detection using directional and non-directional classification. These binary maps represent an overlay of voxels which were declared significant only when using a directional analysis (red), a non-directional analysis (green), or both (blue).

*Discussion:*

We show that in the transition from univariate to multivariate group analysis, the field underwent a paradigm shift in the definition of the null hypothesis. Most MVPA studies to date implicitly employ a non-directional analysis. This means that subjects do not necessarily share the same spatial pattern of BOLD activity and may even have opposite activity patterns. However, the motivation of the first papers that popularized MVPA was that it would allow researchers to discover patterns at lower than single voxel resolution - such as orientation columns in visual cortex (Kamitani and Tong, 2005), or discover weak, subthreshold effects (Haxby, 2012). The bias towards non-directional effects in MVPA analysis is in stark contrast to the hypothesis underlying univariate group analysis - namely that subjects share the same direction (sign) of activation. Indeed we find that employing a directional MVPA analysis uncovers regions that only partially overlap with non-directional analysis, but more importantly also new regions. One may expect regions detected with a directional test to be a subset of regions detected with a non-directional test. This, however, is not the case for the same reason a rejection with a single sided t-test does not imply a rejection with a two sided t-test.

The early papers that popularized MVPA did not make any explicit hypothesis with regard to shared spatial patterns across subjects. Perhaps if these early studies used not only non-directional but also directional tests they would uncover similar representations across subjects. Indeed, we find strong spatial similarities over subjects along the superior temporal gyrus using both the directional and non-directional analysis. The fact that directional and non-directional analysis reveal only partially overlapping brain areas may also help shed light on the poor correspondence that has sometimes been observed between multivariate and univariate group analysis (Jimura and Poldrack, 2012). Importantly, performing a directional or non-directional group analysis does not require the acquisition of new data and both analysis approaches can be employed using the same data set.

Some of the differences we observe between directional and nondirectional tests may be explained by differences in the quality of normalization between directional and non-diretional regions. However, it is unlikely to be the full story. First, the vast literature on univiariate effects has shown that normalization is quite succesfful in achieveing voxel level allignment across subjects. Second, at least in this dataset, the strongest effects in both the directional and non-directional tests are observed in regions in which the signal is directional. Though directional effects may exists even in non-directional regions (obscured by imperfect normzalition) finding directionally responding multivariate regions has important biological implications since it is a positive rather than a negative result.

By proposing an informed choice between directional and non-directional tests, we believe the neuroimaging community will gain a better understanding of the type of signal one is discovering. A sharper definition of the nature of spatial patterns of activity across subjects can have important implications for the study of patient populations and design of brain computer interfaces. For example, in some decoding applications, learning a model which works for all subjects may be desirable. Such a scheme is expected to be fruitful only in regions showing a *directional* signal – or a shared spatial pattern of information across subjects. In contrast, the spatial activity patterns in regions with non-directional signal (i.e. each subject has a unique spatial pattern of information) are expected to generalize to a lesser extent from one subject to another.

Our conclusions are not an artefact of using a more powerful testing approach as the partial overlap between directional/non-directional signal is also present when using a classification based approach. Testing based approaches also have several other advantages on classification based approaches – such as faster run times, avoiding the split into train and test sets and no additional parameters to set. Given these advantages, using our testing based approach can also help augment classification based decoding used in BCI applications by rapidly identifying brain regions most likely to show similar activity patterns across subjects.

Unique spatial patterns across subjects imply poor similarity, and thus low $T^D$ values. Indeed we find high $T^D$ values in the vocal voice areas which this task localizes (Fig. 4). These voice areas were only visible in the original data-set, at the group level using the full set of 218 subjects and computing a probability map of activity. Here, we are able to localize these human voice areas using only a random subset of 20 subjects.

Computing $T^D$ can prove informative in many cases. For example, it is possible that primary sensory regions show similar spatial patterns of activity across subjects (e.g. due to tonotopic representations), whereas higher level association areas are more idiosyncratic and display specific activity patterns unique to each subject. Moreover, once a set of brain regions is discovered to be associated with a certain task, one can probe sub regions using $T^D$ to characterize degrees of multivariate pattern "personalization" across subjects as they relate to hierarchical models of neural processing.

Last, using $T^D$ could aid in targeting regions for brain computer interface (BCI) development. Since one of the challenges of modern BCI implementation is the need to learn a specific classifier for each subject, using $T^D$ can help identify regions which have stable multivariate patterns across subjects so that the same model (classification model –such as SVM) could be used across subjects.

A natural extension of this work would be to assess the replicability of multivariate signals across studies in both directional and non-directional analysis frames. For instance, the different definitions of signals in directional and non-directional hypothesis may also have differential replicability prospects on both the single subject and study level.

**Acknowledgments:**

The study was supported by the I-CORE Program of the Planning and Budgeting Committee and The Israel Science Foundation (grant No. 51/11), Human Frontiers Science Project Organization (HFSPO) (CDA00078/2011-C) and Israel Science Foundation (grants No. 1771/13 and 2043/13) to R.M. and the National Science Foundation (OCI-1131441) to R.A.P.

\* We offer code at https://github.com/roeegilron/Multi-TandFuA and maps at http://neurovault.org/collections/978/ , Original data can be found https://openfmri.org/dataset/ds000158.